\documentclass{aastex}
\usepackage{emulateapj5}
\input epsfig.sty

\begin{document}

\title{Complex Phase Lag Behaviors of the 0.5 -- 10 Hz QPOs in GRS 1915+105}
\author{ D. Lin\altaffilmark{1, 2}, I. A. Smith\altaffilmark{1},
E. P. Liang\altaffilmark{1}, M. B\"ottcher\altaffilmark{1, 3}}

\altaffiltext{1}{Department of Space Physics and Astronomy,
Rice University, 6100 S. Main, Houston, TX 77005, USA.
Email: lin@spacsun.rice.edu.}
\altaffiltext{2}{Currently at Veritas DGC Inc.}
\altaffiltext{3}{Chandra Fellow.}

\begin{abstract}  
Through studying the hard lags between the soft (3.3 -- 5.8 keV)
and hard (13.0 -- 41.0 keV) photons of the 0.5 -- 10 Hz QPOs in
GRS 1915+105, we have classified them into three types: 0.5 -- 2.0
Hz QPOs, 2.0 -- 4.5 Hz QPOs, and 4.5 -- 10 Hz QPOs. They
are closely related to different temporal and spectral states. 
The first type of QPOs (0.5 -- 2 Hz) have positive hard lags at both the QPO 
fundamental and first harmonic frequencies. These QPOs were 
observed in the quiescent soft state. The second  type of QPOs 
(2 -- 4.5 Hz), which were also detected in the quiescent
soft state, have opposite signs of hard lags at the QPO fundamental
and first harmonic frequencies. The third type (4.5 -- 10 Hz), which showed up 
in medium soft quiescent/out-burst states, do not have significant 
higher harmonic peaks. There is a smooth transition
between these three types of QPO behaviors. We did not detect 0.5 -- 10 Hz QPOs
in the very soft state. We discuss some of the 
implications of these results.
\end{abstract}


\section{Introduction}
The superluminal source GRS 1915+105 displays a rich diversity of 
lightcurve morphology, power density spectrum (PDS), quasi-periodic 
oscillations (QPOs), phase lags, and coherence. The lightcurve morphology
varies from out-burst to intermittent out-burst, to quiescence (e.g. 
Belloni et al. 1997, 2000). The PDS shape can be either a broken
power law with a flat top or a simple power law 
(Morgan, Remillard, \& Greiner 1997). 
The QPO fundamental frequency ranges from mHz to 67 Hz 
(e.g. Morgan, Remillard, \& Greiner 1997), and some QPOs were 
detected up to the third harmonic (Cui 1999). 

Several studies have related the temporal behaviors with the 
spectral properties. For example, using the standard disk blackbody and 
power law model, Muno, Morgan, \& Remillard (1999) 
found that the 0.5 -- 10 Hz QPO frequency increases as the 
inner-disk color temperature increases from 0.7 to 1.5 keV. 
Chen, Swank, \& Taam (1997) 
showed that the hardness ratio between the energy bands 11 -- 30.5 keV 
and 2 -- 11 keV is a good indicator for the presence of two types 
of QPOs. When this hardness ratio is above 0.1, GRS 1915+105 exhibited
narrow 0.5 - 6 Hz QPOs. Broad QPOs or no QPO were 
observed when this hardness ratio is below 0.1. According to Markwardt, 
Swank, \& Taam (1999), the 1 -- 15 Hz QPOs were present when the power-law 
component became hard and intense, but the QPO frequency is 
correlated with the parameters of the thermal component. In these studies, 
the 0.5 -- 10 Hz QPOs were treated as a single type of QPOs. However, our 
studies of the phase lag behaviors of the 0.5 -- 10 Hz QPOs
indicate that there are three types of 0.5 -- 10 Hz QPO behaviors
which can be classified according to the QPO frequency: 0.5 -- 2.0 
Hz, 2.0 -- 4.5 Hz, and 4.5 -- 10 Hz.

Complex phase lag (or hard lag) behaviors have been observed in 
the 3 -- 12 Hz QPOs of XTE J1550--564 (Wijnands, Homan, 
\& van der Klis 1999) and in the 66.8 mHz QPO of GRS 1915+105 
(Cui 1999). Wijnands, Homan, \& van 
der Klis (1999) found two types of 3 -- 12 Hz QPOs in XTE J1550--564.
The first type of QPOs have a broad QPO peak with a QPO
frequency of 6 Hz. The hard lags of this type of QPOs are negative
at both the fundamental and first harmonic frequencies.
 The second type of QPOs, 
which have narrow peaks and a fundamental peak frequency of 3 Hz, 
switch signs for the hard lags at the different harmonic frequencies. 
Similarly, Cui (1999) found that the hard lags of the 66.8 mHz QPO
in GRS 1915+105 alternate from negative to positive values as the frequency 
increases from the fundamental to higher harmonic frequencies.

In this letter, we show the different phase lag behaviors 
of 0.5 -- 10 Hz QPOs in 20 {\it Rossi X-ray Timing Explorer (RXTE)}
observations of GRS 1915+105. We find that the phase lag behaviors 
are closely related to the QPO fundamental frequency and the spectral states.
In section 2, we describe how the data were analyzed. In section 3,
the analysis results are presented. In the last section, we  
discuss the implications of these results.

\section{Data Reduction}

To examine the phase lag behaviors, we selected the $ RXTE $ observations 
published in Morgan, Remillard, \& Greiner (1997) that showed QPOs in 
the frequency range of 0.5 -- 10 Hz. We also picked several other 
observations in late 1996 and 1997 which cover a big dip in the $RXTE$
All Sky Monitor (ASM) light curve of GRS 1915+105. A full list of the 
observations is given in Table 1.
The radio fluxes at the times of some of these observations are given in 
Pooley \& Fender (1997) and Fender et al. (1999).

We extracted light curves using 
three energy bands, 3.3 -- 5.8 keV, 5.8 -- 13.1 keV, and 13.1 -- 41.0 keV. 
The channels in the energy range of 2.5 -- 3.3 keV were not used,
because they have been binned together with the channels below 2.5 keV
in the archival high time resolution data. The counts were summed into 
time bins of 7.8125 ms, and timing analyses were performed on 
intervals of 256 seconds. 

The timing analyses included calculations
of power density spectra and cross-spectra. In order
to attenuate noise, we averaged the power spectra and 
cross-spectra over all observation intervals. 
In displaying the power density spectra, we 
used the Leahy normalization (Leahy et al. 1983) without 
subtracting white noise. Phase lags and coherence between
signals at two different energy channels were obtained from the 
cross-spectra. We used Monte Carlo simulations to estimate the
phase lag errors from the cross spectra. 
In the simulations, we assumed that the values
of the real and imaginary parts had normal distributions, and used 
the fact that the  cross-spectrum is a linear function of detector 
count rates.

We classified the GRS 1915+105 lightcurve profiles into two
general states. One is the ``quiescent'' state in which the count rates on the 
1-s time resolution lightcurves vary by less than 20\% of the 
average count rates and there are no spikes wider than
10 seconds. The other is the ``out-burst'' or ``flaring'' state which has spikes
wider than 10 seconds and count rates varying by more than
20\% of the average value, but does not have large scale structures
in the light curves. Belloni et al. (2000) classified the observations
of GRS1915+105 into 12 classes (hereafter Belloni classes), 
based on their count rate and hardness characteristicsi.
Our ``quiescent'' state  corresponds to Belloni classes $\phi$ and $\chi$,
and the ``flaring'' state includes classes $\gamma$, $\mu$, $\delta$
and $\beta$.

We used a simple power law model plus interstellar absorption to fit the photon 
spectra in the energy range of 2.5 -- 40 keV. 
Though this simple model does not give
good fits to the data, the photon power-law index $\alpha$ is a good indicator
of the spectral hardness. We defined three spectral states: soft state 
($\alpha$ = 2.0 -- 3.0), medium soft state ($\alpha$ = 3.0 -- 4.0), 
and very soft state ($\alpha > 4.0$).

\begin{figure*}[t]
\epsfig{figure=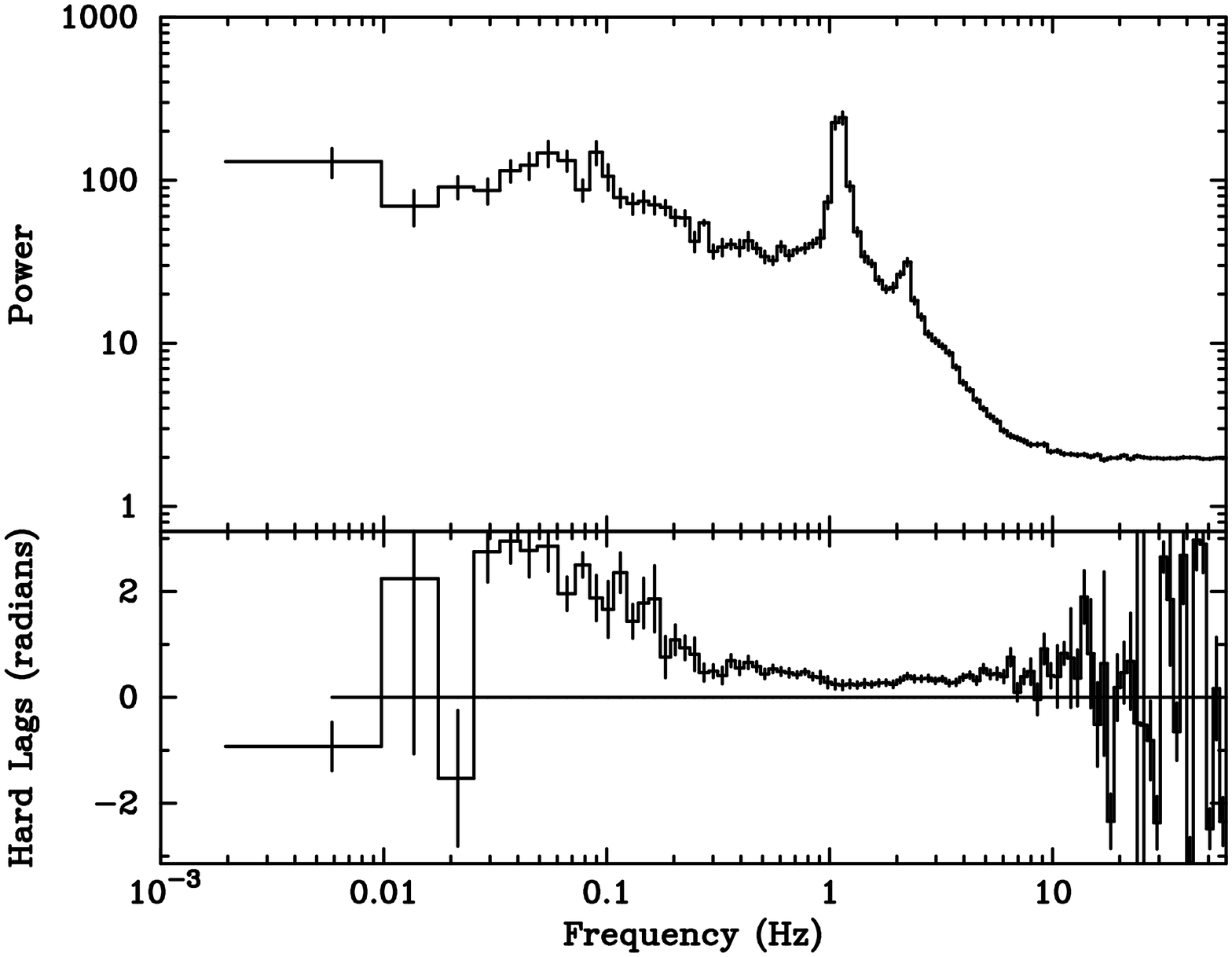,width=2.3truein,angle=270}
~
\epsfig{figure=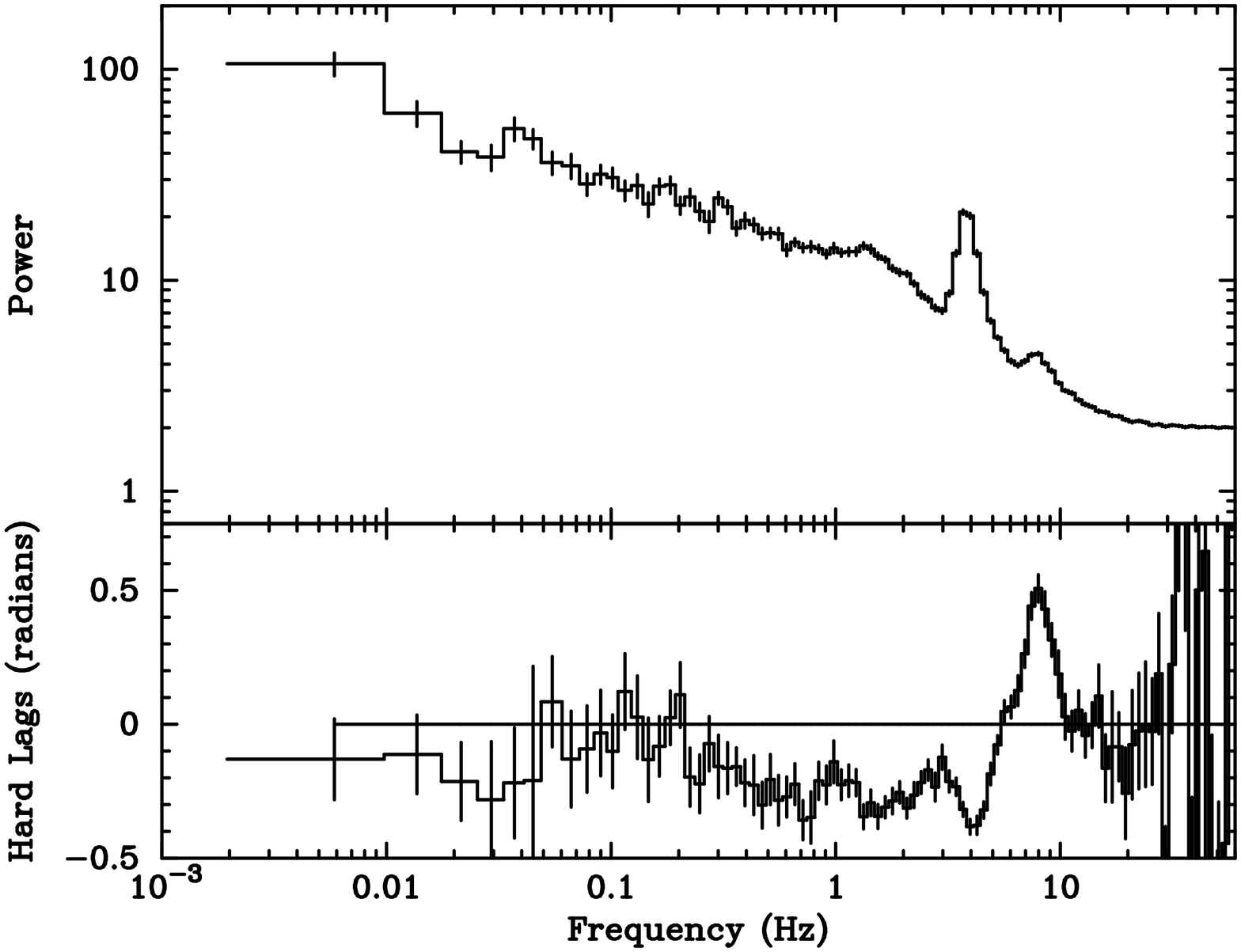,width=2.3truein,angle=270}
~
\epsfig{figure=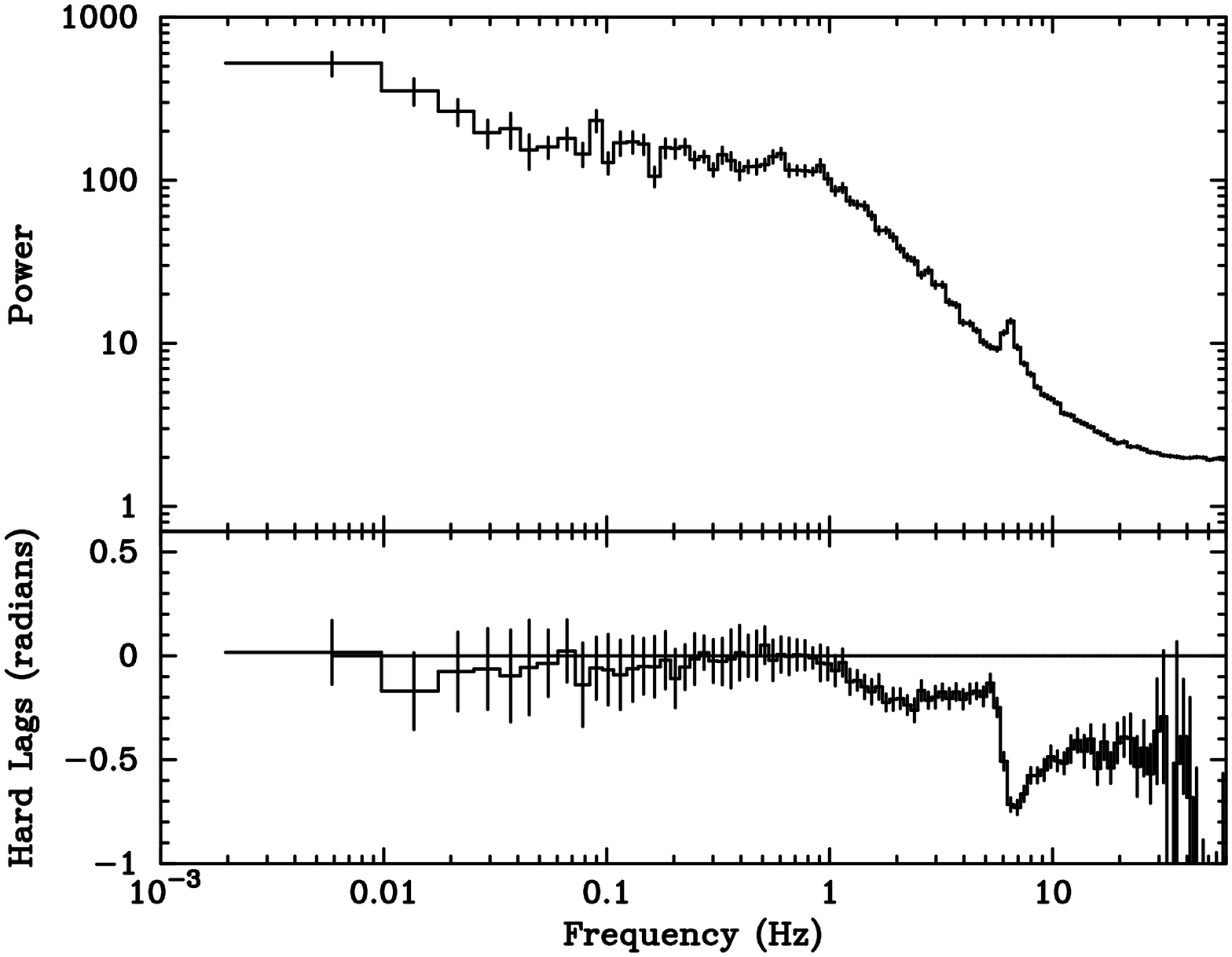,width=2.3truein,angle=270}
\figcaption[fig1.eps]{The Leahy power density spectrum and hard lags for an 
example of the 0.5 -- 2.0 Hz QPOs. The observation was made on July 19th, 1996. 
In this figure and the following Figures 2 \& 3, the upper panel is 
the power density spectrum
of the energy band 3.3 -- 5.8 keV, and the lower panel is the hard
lags between the energy bands 3.3 -- 5.8 keV and 13.1 - 41.0 keV. 
The QPO fundamental frequency is 1.14 Hz. The hard lags are positive,
i.e. the hard photons arrive after the soft photons.}
\figcaption[fig2.eps]{The Leahy power density spectrum and hard lags for an 
example of the 2.0 -- 4.5 Hz QPOs. 
The observation was made on December 24th, 1996.  
The QPO fundamental frequency is 3.69 Hz. 
The hard lag  is negative at the fundamental 
frequency but positive at the first harmonic frequency.}
\figcaption[fig3.eps]{The Leahy power density spectrum and hard lags for 
an example of the 4.5 -- 10 Hz QPOs. The observation was made on 
August 31st, 1996. 
The QPO fundamental frequency is 6.47 Hz. The hard lag 
 turns further negative at the fundamental frequency. No significant first 
harmonic peak is detected.}
\end{figure*}

\section{Analysis Results}
The results are summarized in Table 1. The observations
are listed in the increasing order of the QPO fundamental frequency.
Three types of hard lag behaviors were found in the observations.
The hard lags at the fundamental and first harmonic frequencies
are both positive when the fundamental frequency is low. When the 
fundamental frequency increases above 2 Hz but below
4.5 Hz, the hard lags become negative at the fundamental frequency 
but stay positive at the first harmonic frequency. 
When the fundamental frequency moves to even
higher values, no significant first harmonic peak can be detected.
While there is a smooth transition between these behaviors,
to highlight their differences we describe the three types
separately in the following sub-sections.

\subsection{0.5 -- 2 Hz QPOs}

Figure 1 shows an example of the 0.5 -- 2 Hz QPOs and the hard 
lags between the two energy bands (3.3 -- 5.8 keV and 13.1 -- 41.0 keV).
Both the fundamental and first harmonic peaks are significantly detected. 
The hard lags at both peaks are 
positive and consistent with the values at the surrounding continuum
frequencies. Among the four observations that detected the 
0.5 - 2 Hz QPOs, the coherence between the two energy bands
at the QPO fundamental frequency is
0.95 or 0.96, which is also consistent with the overall distribution of 
the coherence. The photon spectrum power-law index is in the range of
2.6 -- 2.7, which indicates that GRS 1915+105 was in the soft state.
As the QPO fundamental frequency increases, the hard lag at the 
fundamental frequency decreases while the the hard lag at 
the first harmonic frequency increases. The X-ray  flux in the range 
of 3.3 -- 13.1 keV was essentially the same among the four observations, 
and the source was in the ``quiescent'' state.

\subsection{2.0 -- 4.5 Hz QPOs}
The 2.0 -- 4.5 Hz QPOs have very peculiar
phase lag behaviors (for example, see Figure 2). The hard lags are 
negative at the QPO fundamental frequency but turn positive at the 
first harmonic frequency. They significantly deviate from the 
overall trend in the hard lag distribution over frequencies. 
The photon power-law index is in the range of 2.5 -- 3.0. Thus GRS 1915+105
was in the soft and quiescent state. Though the hard lags at the fundamental
frequency appear to generally become more negative at higher 
fundamental frequencies, this trend is not as well defined as that for the 
0.5 -- 2.0 Hz QPOs.
Among the nine observations that
detected the 2 - 4.5 Hz QPOs, the total 3.3 -- 13.1 keV X-ray flux varied by a 
factor of more than three, and had no apparent correlations with the
QPO fundamental frequency. However, spectral analyses by 
Trudolyubov, Churazov, \& Gilfanov (1999) showed that the fundamental
frequency positively correlates with the flux of the 
soft blackbody component that was obtained by fitting the spectrum with
the standard disk backbody plus an exponentially cutoff power law.
 
The coherence is generally high at the fundamental frequency (above 0.85)
except for the observation on July 5th, 1997, which may be treated
as a special case because the analysis was done on the quiescent
segments of an intermittent out-burst state. 
The coherences in the intermittent out-burst state were significantly 
lower than the coherences in both the quiescent and flaring states.
The implications of this need to be further investigated.

\subsection{4.5 -- 10.0 Hz QPOs}
When the QPO frequency is above 4.5 Hz, the first harmonic
QPO peak can not be significantly detected, and we do not see
a change in the sign of the hard lags between the fundamental
and where the first harmonic frequencies would be (for example, see Figure 3). 
The hard lags 
at both frequencies are negative. The coherence at the fundamental
frequency between the two energy bands (3.3 -- 5.8 keV
and 13.1 -- 41.0 keV) is lower than for the
previous two types of QPOs. The source was in the medium 
soft state with a photon spectrum power law index of 3.1 -- 3.4. 
Unlike the previous two types of QPO, 4.5 -- 10 Hz
QPOs can be present in both quiescent and out-bursting states.
The 3.3 -- 13.1 keV X-ray flux is also very diverse. 

\subsection{Epochs without 0.5 -- 10.0 Hz QPOs}
We have three observations that did not 
detect 0.5 -- 10.0 Hz QPOs. One common feature among these
observations is that the source was in the very soft state
with a photon power-law index above 4.0 (Table 1). 
Similar results were also reported by Muno, Morgan, \& Remillard (1999).
Among the three observations, one has a PCA 3.3 -- 13.1 keV count rate
of 6889 counts/s, compared to 16800 and 24005 counts/s in the other 
two observations. Therefore, the source was not necessarily 
in the very high state even though the spectra were very soft.

\section{Discussion}
Through studying the hard lag behaviors of the 0.5 -- 10 Hz QPOs,
we have highlighted them into three types that
are closely related to different temporal and spectral properties. 
The first type of QPOs (0.5 -- 2 Hz) have positive hard lags at 
the fundamental and first harmonic frequencies and were 
observed in the quiescent soft state.
The second type of QPOs (2 -- 4.5 Hz) were also detected in the quiescent
soft state but have opposite signs for the hard lags at the fundamental
and first harmonic frequencies.  The third type (4.5 - 10 Hz), which 
showed up in medium soft quiescent/out-burst states, do not have significant 
harmonic peaks. There is a general trend that the hard lags at the QPO 
fundamental frequency
decrease from positive to negative values as the fundamental frequency
moves from 0.66 Hz to higher values. We also found that the photon
spectra become softer as the QPO fundamental frequency increases.
After we presented this work to the 195th AAS meeting in Atlanta
on January 14th, 2000, it was brought to our attention that
a recent paper by Reig et al. (2000) also reported such
trends in GRS 1915+105.

The opposite signs of the hard lags at the QPO 
fundamental and first harmonic frequencies do 
not necessarily mean that the photon arrival times
are different at different QPO harmonic frequencies.
For example, the Fourier transform of a decaying oscillating signal
$f (t) = e^{-\lambda t} (cos(\omega_0 t) + cos(2 \omega_0 t)) $ is:
\begin{eqnarray}
  F(\omega) & = & \frac{1}{\lambda - i(\omega-\omega_0)} + \frac{1}
  {\lambda - i(\omega+\omega_0)} \nonumber \\
            &   & + \frac{1}{\lambda - i(\omega-2\omega_0)} 
 + \frac{1}{\lambda - i(\omega + 2\omega_0)}
\end{eqnarray}
The power density has two QPO peaks at $\omega_0$ and $ 2\omega_0$
respectively.
For $\omega > 0$, the two dominant terms are:
\begin{equation}
  F(\omega) \approx \frac{1}{\lambda - i(\omega-\omega_0)}  + \frac{1}{\lambda - 
i(\omega-2\omega_0)} 
 \end{equation} 
So we have $F(\omega=\omega_0) \approx \frac{1}{\lambda}  + \frac{1}{\lambda + 
i(\omega_0)}$
and  $F(\omega=2 \omega_0) \approx \frac{1}{\lambda}  + \frac{1}{\lambda - 
i(\omega_0)}$.
Therefore, $F(\omega=\omega_0) \approx F(\omega=2 \omega_0)^* $, i.e. they have
opposite signs for the phase term even though the oscillations have no time 
delays from each other. The two additional minor terms make the conclusion less 
obvious, but we still can find appropriate $\lambda_0$ and $\omega_0$ to 
make the phase lags have opposite signs. Therefore, it may be misleading
to interpret the phase lag at a particular QPO frequency as simply the time 
delay between two oscillating signals. 

In the classic multi-color thermal disk 
models, the negative hard lags can be explained by assuming that 
perturbations propagate from the 
inner disk to the outer disk. The perturbations on the inner edge may come from
the central object or from the magnetic fields that are being sucked into
the central object. Zhang, Cui, \& Chen (1997) and Cui, Zhang, \& Chen (1998)  
suggested that GRS 1915+105
contains a black hole rapidly spinning in the same direction as the accretion
disk. In such a prograde system, the last stable orbit is much smaller than
the non-spinning system. Therefore, strong perturbations on the inner edge
of the accretion disk are very likely. There may also be perturbations 
propagating inwards from the outer edge of the disk, which generate positive 
hard lags. When the disk inner edge is farther 
away from the central object and thus the QPO frequency is smaller, it is 
natural  to assume that the perturbations on the inner edge is weaker. 
Therefore, the inward perturbations would dominate over the outward 
perturbations, and the hard lags are expected to be positive. 
This could explain why the lowest 0.5 -- 2 Hz QPOs have positive hard 
lags while the faster ones have negative hard lags.  

Taam, Chen, \& Swank (1997) proposed that the out-bursts in GRS 1915+105 are 
due to the ejection of the inner disk. A direct prediction of 
this model is that the inner disk edge is farther
away from the central object in the out-burst state than in the quiescent
state.  Our results may pose problems to this prediction. The 
observation made on July 5th, 1997, for example, had long out-burst and  
quiescent segments. We therefore performed separate timing analyses 
on the out-burst and quiescent segments. 
We found that the quiescent state has a lower QPO 
fundamental frequency than the out-burst state, and thus the disk 
inner edge should be closer to the central object in the out-burst 
state than in the quiescent state if we assume the QPO fundamental 
frequency is somehow related to the sound crossing time across
the disk inner edge. 
   
\acknowledgements{We thank S. Nan Zhang for his helpful comments.
This work was partially supported by NASA grant NAG 5-3824, and
made use of the {\it RXTE} ASM data products provided by the ASM/{\it RXTE}
teams at MIT and by the {\it RXTE} SOF and GOF at the NASA's Goddard Space 
Flight Center. 
The work of MB is supported by NASA through Chandra Postdoctoral Fellowship 
grant PF 9-10007 awarded by the Chandra X-ray Center, which is operated by 
Smithsonian Astrophysical Observatory for NASA under contract NAS 8-39073.}
                                                        

%
%

\renewcommand{\baselinestretch}{0.8}
\begin{deluxetable}{lllllllll}
\tablecaption{QPO properties for each observation \tablenotemark{a}}
\tablehead{
\colhead{$f_0$ \tablenotemark{b}} & \colhead{$f_1$\tablenotemark{c}} & 
\colhead{$\phi_0$ \tablenotemark{d}} & \colhead{$\phi_1$ \tablenotemark{e}} & 
\colhead{Cohn \tablenotemark{f}} & \colhead{$\alpha$ \tablenotemark{g}} & 
\colhead{PCA \tablenotemark{h}} & \colhead{F/Q \tablenotemark{i}} & 
\colhead{Date} }
\tablewidth{0pt}
\scriptsize
\startdata 
$ 0.648 \pm 0.006$ & Y & $ 0.40 \pm 0.07$    & $0.27 \pm 0.03$ & 0.95 & 2.6 & 
$9888$ & Q & 1996-07-26 \\
$ 1.000 \pm 0.004$ & Y & $ 0.27 \pm 0.03$    & $0.36 \pm 0.02$ & 0.96 & 2.6 & 
$9713$ & Q & 1996-08-03 \\
$ 1.112 \pm 0.004$ & Y & $ 0.24 \pm 0.02$    & $0.38 \pm 0.03$ & 0.95 & 2.7 & 
$10336$ & Q & 1996-07-19 \\
$ 1.683 \pm 0.008$ & Y & $ 0.09 \pm 0.01$  & $0.43 \pm 0.03$ & 0.95 & 2.7 & 
$9940$ & Q & 1996-08-10 \\
\hline
$ 2.263 \pm 0.004$ & Y & $ -0.05 \pm 0.01$ & $0.52 \pm 0.04$  & 0.95 & 2.9 & 
$11107$ & Q & 1996-07-16 \\
$ 2.708 \pm 0.007$ & Y & $ -0.13 \pm 0.01$ & $0.48 \pm 0.04$  & 0.92 & 2.8 & 
$7212$ & Q & 1997-10-08 \\
$ 2.970 \pm 0.010$   & Y & $ -0.22 \pm  0.02 $ & $0.58 \pm 0.03 $ & 0.88 & 
2.6 & $3137$ & Q & 1997-02-22 \\
$ 3.241 \pm 0.008$ & Y & $ -0.19 \pm 0.01$ & $0.33 \pm 0.02$ & 0.88 & 2.6 & 
$7136$ & Q & 1996-10-29 \\
$ 3.370 \pm 0.020$   & Y & $ -0.21 \pm 0.08$   & $0.31 \pm 0.08$  & 0.19 & 
2.6 & $3640$ & Q & 1997-07-05 
\tablenotemark{j} \\
$ 3.475 \pm 0.008$ & Y & $  -0.27 \pm 0.02 $ & $ 0.48 \pm 0.19$   & 0.93 & 3.0 
& $12174$ & Q & 1996-07-11 \\
$ 3.507 \pm 0.008$ & Y & $ -0.27 \pm 0.01$ & $  0.33 \pm 0.11$  & 0.92 & 3.0 & 
$12101$ & Q & 1996-07-14 \\
$ 3.860 \pm 0.009$ & Y & $ -0.36 \pm 0.02$ & $0.47 \pm 0.03$    & $0.86 $ & 2.6 
& $5212$ & Q & 1996-12-24 \\
$ 4.088 \pm 0.007$ & Y & $ -0.30 \pm 0.02  $ & $0.15 \pm 0.06$    & 0.91 & 2.9
& $13325$ & Q & 1996-08-25 \\
\tableline
$ 4.54 \pm 0.01$   & N & $ -0.38 \pm 0.02 $  & N/A  & 0.88 & 3.1 & $14302$& F & 
1996-08-18 \\
$ 6.09 \pm 0.07$ \tablenotemark{k} & N & $ -0.34 \pm 0.04 $ & N/A  & 0.55 & 3.4
& $11987$ & F & 1997-07-07 \\
$ 6.45 \pm 0.03$   & N  & $-0.71 \pm 0.03$    & N/A  & 0.65 & 3.2 & $24510$ & Q 
& 1996-08-31 \\
$ 6.90 \pm 0.10$   & N  & $ -0.20 \pm 0.03$ & N/A  & 0.12 & 3.1 & $6942$ & F & 
1997-07-05 
\tablenotemark{j} \\
$ 7.44 \pm 0.04$   & N  & $ -0.56\pm 0.03$ & N/A  & 0.68 & 3.4 & $14417$  & F & 
1996-10-15 \\
\tableline
No   & N  & N/A   & N/A  & N/A  & 4.4 & $6889$  & F & 1996-06-29 \\
No   & N  & N/A   & N/A  & N/A  & 4.0 & $16800$  & Q & 1997-07-20 \\
No   & N  & N/A   & N/A  & N/A  & 4.0 & $24005$  & F & 1997-12-22 \\
\enddata
\tablenotetext{a}{The hard lags and coherences were calculated between
the energy bands 3.3 -- 5.8 keV and 13.1 -- 41.0 keV.}
\tablenotetext{b}{The QPO fundamental frequency (Hz).}
\tablenotetext{c}{Whether the first harmonic peak is present or not, `Y' for
yes, and `N' for no.}
\tablenotetext{d}{The hard lag at the fundamental frequency (in radians).
`N/A' means not applicable because no QPO is detected.}
\tablenotetext{e}{The hard lag at the first harmonic frequency (in radians).
`N/A' means not applicable because no first harmonic peak is detected.}
\tablenotetext{f}{The coherence value at the fundamental frequency. 
`N/A' means not applicable because no QPO is detected.}
\tablenotetext{g}{The photon spectrum power-law index.}
\tablenotetext{h}{The average PCA count rate in the range of 3.3 -- 13.1 keV 
(Counts/s).}
\tablenotetext{i}{Whether the source is in the flaring state `F'
or in the quiescent state `Q'.}
\tablenotetext{j}{The observation was split into quiescent segments and
out-burst segments. The timing analysis has been done separately on each 
type of segment.}
\tablenotetext{k}{A broad QPO peak.}
\end{deluxetable}

\end{document}